\documentclass[12pt,a4paper,keywords,prl,groupedaddress,showkeys]{revtex4}
\usepackage{epsfig}
\usepackage{setspace}
\usepackage{graphicx}% Include figure files
\usepackage{color}
\usepackage{amsmath}
\usepackage{epstopdf}
\usepackage{amsbsy}
\usepackage{dcolumn}% Align table columns on decimal point
\usepackage{bm}% bold math

\newcommand{\vE}{\mbox{\boldmath$E$}}

\newcommand{\vH}{\mbox{\boldmath$H$}}

\newcommand{\vP}{\mbox{\boldmath$P$}}

\begin{document}

\title{Bi-quadratic magnetoelectric coupling in underdoped La$_2$CuO$_{4+x}$}
\author{S. Mukherjee$^{1,2,3}$}
\author{B. M. Andersen$^2$, Z. Viskadourakis$^1$, I. Radulov$^1$, C. Panagopoulos$^{1,4,5}$}
\affiliation{
$^1$ Institute of Electronic Structure and Laser, Foundation for Research and Technology Hellas, Heraklion, 70013, Greece\\
$^2$Niels Bohr Institute, University of Copenhagen, DK-2100 Copenhagen \O, Denmark, email:shantanu@fys.ku.dk, phone:+45 35325504, fax:+45 35320460\\
$^3$Niels Bohr International Academy, Niels Bohr Institute, University of Copenhagen, Blegdamsvej 17, DK-2100 Copenhagen \O, Denmark\\
$^4$Department of Physics, University of Crete, Heraklion, 71003, Greece\\
$^5$ Division of Physics and Applied Physics, Nanyang Technological University, 637371, Singapore
}

%\date{\today{}}

\begin{abstract}
The recent discovery of relaxor ferroelectricity and magnetoelectric effect in lightly doped cuprate material La$_2$CuO$_{4+x}$ has provided a number of questions concerning its theoretical description. It has been argued using a Ginzburg-Landau free energy approach that the magnetoelectric effect can be explained by the presence of bi-quadratic interaction terms in the free energy. Here, by using the same free energy functional, we study the variety of behavior which can emerge in the electric polarization under an external magnetic field. Subsequently, we discuss the role of Dzyaloshinskii-Moriya interaction in generating this magnetoelectric response. This work is particularly relevant for such relaxor systems where the material-dependent parameters would be affected by changes in e.g. chemical doping or cooling rate.

\end{abstract}


\maketitle

\section{Introduction}

The parent cuprate materials are Mott insulating antiferromagnets that display a wide variety of novel ground states upon doping, including glassy magnetic phases and unconventional superconductivity. However, a detailed understanding of how the dopants lead to such complex physics is yet to be obtained. Some open issues are related to the structural properties of the dopant atoms such as their preferred locations within the host lattice, the physics of e.g. oxygen ions to form clusters, and also the dynamical properties of dopants. One interesting example in this context is the oxygen doped material La$_2$CuO$_{4+x}$ (LCO) which, in the undoped case, contains an antiferromagnetic phase with N\'{e}el temperature T$_N\sim 325K$. Structurally the oxygen dopant ions take spatially non-stoichiometric positions unlike e.g. strontium that acts as a substitutional dopant by replacing lanthanum ions from the host lattice. Experimentally, a small amount of non-stoichiometric oxygen ions will nearly always be present in LCO which, as shown in Refs. \onlinecite{Zach:2011,Mukherjee:2012}, can be utilized to study the unusual physical properties of the very lightly doped cuprates.

We have recently discovered that the nearly undoped La$_2$CuO$_{4+x}$ is a relaxor ferroelectric at low temperatures $T$.\cite{Zach:2011} Further, there exists a weak magnetoelectric effect that is anisotropic for different directions of electric polarization and external magnetic field. In a recent paper, we were able to reproduce the qualitative features of the magnetoelectric curves using a Ginzburg-Landau (GL) analysis that was governed by the effects of a bi-quadratic magnetoelectric coupling.\cite{Mukherjee:2012} Here, we explore the parameter space of the GL free energy to study additional effects in the behavior of $\vP(\vH)$ and in the predicted enhancement of magnetization and magnetocapacitive effect proposed in Ref. \onlinecite{Mukherjee:2012}. It is important to analyze these variations in the coefficients of the magnetoelectric coupling particularly for relaxor systems, since the coefficients of the GL theory are generally material dependent and may be affected by experimental conditions such as e.g. the concentration of dopants, or cooling rate.

\section{Model}

The magnetic structure of LCO is that of a two-dimensional antiferromagnet with weak interplanar exchange coupling giving rise to three-dimensional long-range N\'{e}el order.\cite{Birgeneau:1990,Keimer1:1992,Kastner:1998} The Cu spins are slightly canted out of the CuO$_2$ planes because of a finite Dzyaloshinskii-Moriya (DM) interaction existing in the low temperature orthorhombic phase (LTO).\cite{Thio:1988} On application of an external magnetic field, a first order spin flop transition is observed at a critical magnetic field $H_{\rm{sf}}\sim5\rm{T}$.\cite{Thio:1988,Reehuis:2006} Clear evidence for coupled spin and charge degrees of freedom in these systems come from the observation of pronounced discontinuities in resistivity and dielectric constant at a magnetic field corresponding to $H_{\rm{sf}}$.\cite{Zach:2011,Thio:1990}

Below, we follow the theoretical model introduced in Ref. \onlinecite{Mukherjee:2012} and study the parameter space for the magnetoelectric interaction terms that are responsible for spin-charge coupling in LCO. We further look at the feedback effect on magnetization at low $T$ due to the presence of electric polarization. Below approximately 530K the crystal structure of LCO  is LTO with space group Cmca (D$_{2h}^{18}$). Taking into account the symmetry properties of the Cmca space group, the free energy can be expressed as a sum of three contributions
\begin{align}
F=F_M+F_{MP}+F_P.
\end{align}
Here, $F_M$ is the magnetic free energy, $F_{MP}$ is the magnetoelectric contribution, and $F_{P}$ is the polarization free energy.
The magnetic free energy that accounts for the crystal structure of the LTO phase has been studied previously by e.g. Thio {\it et al.} \cite{Thio:1994} and is given by
\begin{align}
& F_M = {1\over 2}\sum_{i=1}^2[{\chi_{2D}^{-1}\over 2}L_i^2+{1\over 4}AL_i^4+{1\over 6}BL_i^6-CL_iM_i\nonumber\\
 & +{\chi_0^{-1}\over 2}M_i^2 -H_cM_i-H_{ab}L_i]+{1\over 2}J_{\bot}L_1L_2.
\end{align}
Here, the out-of-plane ($c$ direction) [in-plane ($a-b$ plane)] applied magnetic field is represented by $H_c$ [$H_{ab}$]. The coefficients A, B, and C are in general $T$ dependent.\cite{coeff} The order parameter $M_i=(S_{Ai}+S_{Bi})/2$ is the ferromagnetic moment per spin with $S_{Ai},S_{Bi}$ being the sub-lattice spins in the $i^{th}$ plane, and $L_i=(S_{Ai}-S_{Bi})/2$ is the antiferromagnetic order parameter ($L_i||a$). The spins are slightly canted due to the DM interaction term $-CM_iL_i$, which causes them to lie in the $a-c$ plane of the magnetic unit cell. The coupling between the different planes is included by the $J_{\bot}$ term.

The presence of an inversion symmetry in the space group of the crystal forbids any linear magnetoelectric effect \cite{Dzyaloshinskii:1991} and the physics is dominated by non-linear coupling terms. We can focus on the largest non-linear terms by further noting that the experimentally observed polarization response is symmetric under inversion of the external magnetic field (i.e $\vP(\vH)=\vP(-\vH)$). This implies that the dominant couplings are of even order in the magnetic order parameter. Hence, the following terms contribute to the magnetoelectric coupling
\begin{align}
& F_{MP}= \sum_{\alpha,i}({\gamma_{1\alpha}\over 2}L_i^2+{\gamma_{2\alpha}\over 2}M_i^2+\gamma_{3\alpha}M_iL_i)P_\alpha^2,
\end{align}
where the components for $\vP$ run over $\alpha=(a,b,c)$ in the magnetic unit cell.

The polarization free energy is given by
\begin{align}
F_P =  \sum_{\alpha}({\chi_{e\alpha}^{-1}\over 2}P_{\alpha}^2+{\beta\over 4}P_{\alpha}^4) -\vE\vP.
\end{align}
Here, $\chi_{e\alpha}$ is the electric susceptibility for the $\alpha$ component of the polarization, and $\vE$ denotes the applied electric field. The solutions that determine $\vP(\vH)$ are obtained by minimizing $F$ with respect to the electric polarization and magnetic order parameters. In the case of LCO studied experimentally, $T_N\sim 320$K, which is much higher than the $T$ at which the ferroelectric order sets in ($T_P\sim 4.5$K).\cite{Zach:2011} Therefore, we evaluate F$_M$ for the high $T$ phase with $\vP=0$, providing the following set of equations
\begin{align}\label{Meq}
& M_i=\chi_0(H_c+CL_i),\\
& \chi_{2D}^{-1}L_1+AL_1^3+BL_1^5+{1\over 2}J_{\bot}L_2 = CM_1+H_{ab},\\
& \chi_{2D}^{-1}L_2+AL_2^3+BL_2^5+{1\over 2}J_{\bot}L_1 = CM_2+H_{ab},\\
& [\chi_{e\alpha}^{-1} +\sum_{i=1}^2(\gamma_{1\alpha}L_i^2+\gamma_{2\alpha}M_i^2+\gamma_{3\alpha}M_iL_i)]P_\alpha = -\beta P_\alpha^3.
\end{align}

The experimental magnetization curves at low temperatures ($<\sim30$K) have a glassy contribution.\cite{Mukherjee:2012} These features cannot be obtained from the above equations and to include them to lowest order, we take the experimental magnetization values at $T=5$K as input to the model.

\section{Results}

\begin{figure}[t]
\begin{minipage}{.49\columnwidth}
\includegraphics[clip=true,width=0.99\columnwidth]{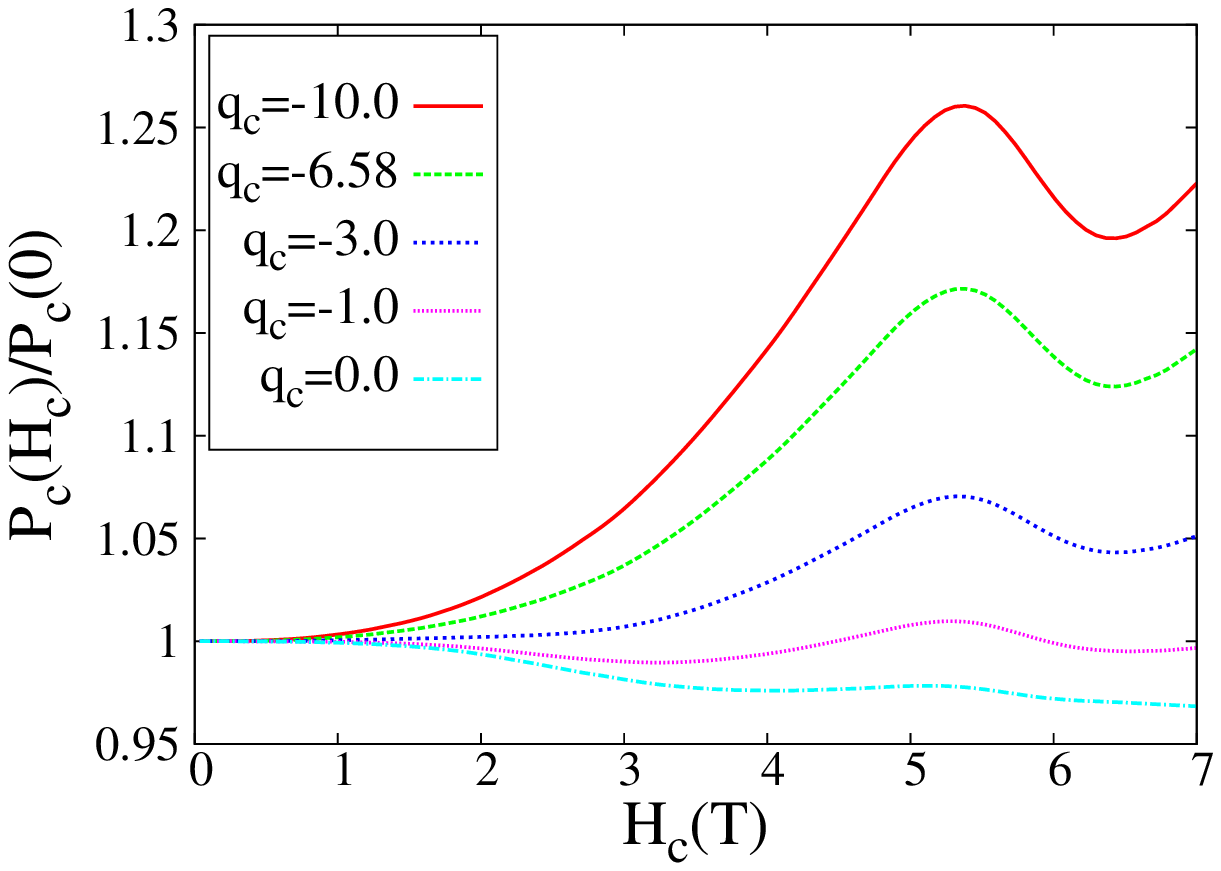}
\end{minipage}
\begin{minipage}{.49\columnwidth}
\includegraphics[clip=true,width=0.99\columnwidth]{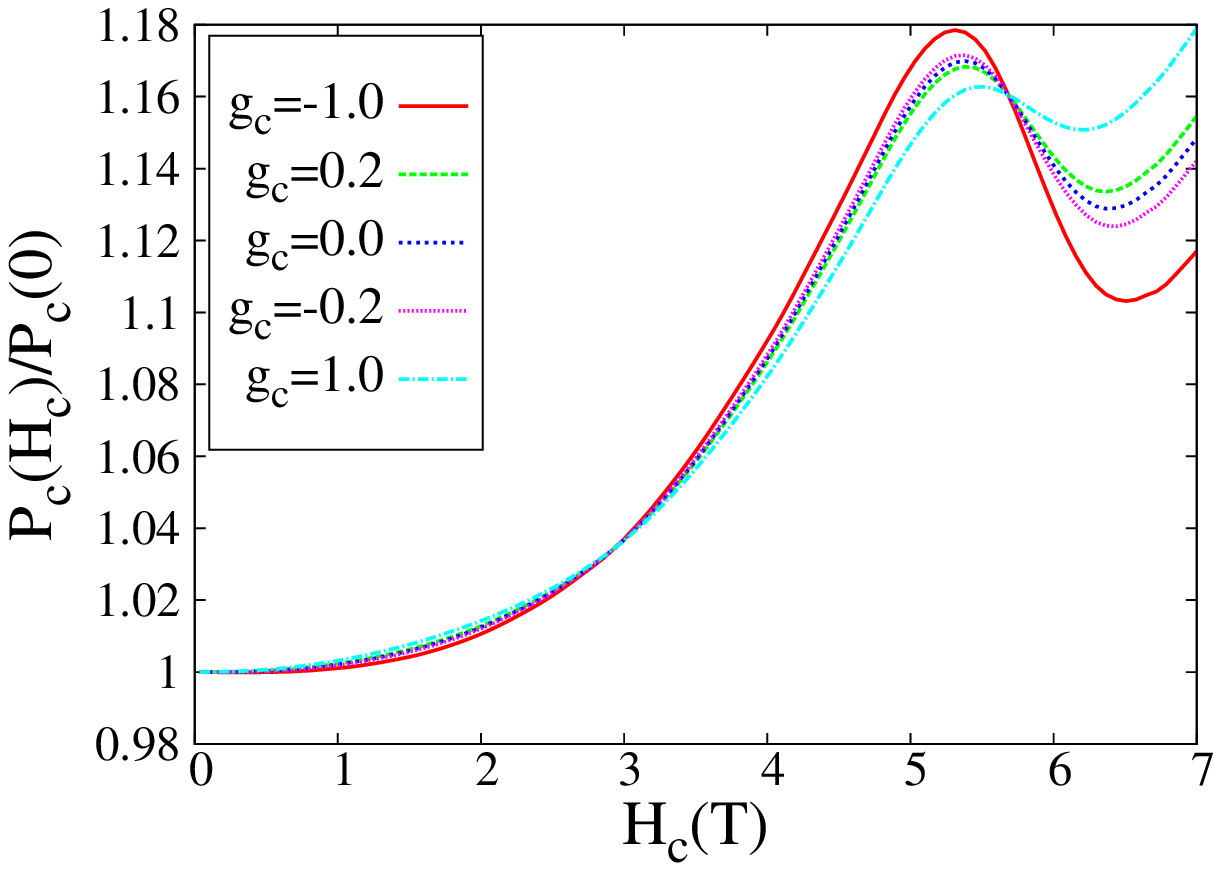}
\end{minipage}
\caption{(Color online) Theoretically calculated $P_{c}(H_{c})/P_{c}(0)$ for different values of $q_c$ and $g_c$. Parameters not mentioned in the plots have been fixed at $s_c=0.1$, $s_a=0.074$, $g_a=g_b=g_c=-0.2$, $q_{c}=-6.58$, and $q_a=q_b=0$ that explain the experimental curves at $T=5\rm{K}$. } \label{fig:theory1}
\end{figure}

In terms of the following rescaled quantities $l_{+}=\chi_0C(L_1+L_2)/2$,
$l_{-}=\chi_0C(L_1-L_2)/2$, $M=(M_1+M_2)/2$, $\gamma_{1\alpha}'=2\gamma_{1\alpha}(\chi_0C)^{-2}$,
$\gamma_{2\alpha}'=2\gamma_{2\alpha}$, $\gamma_{3\alpha}'=2\gamma_{3\alpha}(\chi_0C)^{-1}$,
the polarization dependence on the applied magnetic field can be expressed as\cite{Mukherjee:2012}
\begin{align}\label{Pabeqn}
& {P_{\alpha}(H_{ab})\over P_{\alpha}(0)}=[1+{s_{\alpha}\over l_{-}^2(0)}(l_{-}^2(H_{ab})-l_{-}^2(0)+l_{+}^2(H_{ab}))]^{1/2},\\
& {P_{\alpha}(H_c)\over P_{\alpha}(0)}=[1+{s_{\alpha}\over l_{-}^2(0)}(l_{-}^2(H_c)-l_{-}^2(0)+g_{\alpha} M(H_c)^2\nonumber\\
& +(1-g_{\alpha}-q_{\alpha})l_{+}^2(H_c) +q_{\alpha} M(H_c)l_{+}(H_c))]^{1/2},
\end{align}
where $s_{\alpha}=\lambda_\alpha l_{-}^2(0)/ (\chi_{e\alpha}^{-1}+\lambda_\alpha l_{-}^2(0))$,
$g_\alpha={\gamma_{2\alpha}'/\lambda_{\alpha}}$, and $q_\alpha=\gamma_{3\alpha}'/\lambda_\alpha$ with
$\lambda_\alpha=\gamma_{1\alpha}'+\gamma_{2\alpha}'+\gamma_{3\alpha}'$. In general, all three parameters $s_\alpha$, $g_\alpha$, and $q_{\alpha}$ are $T$ dependent. The $T$ dependence of $s_{\alpha}$ results primarily from its relation to the electric susceptibility.

In the case of an in-plane magnetic field $H_{ab}$, $P_{\alpha}(H_{ab})$ depends on a single fitting parameter $s_{\alpha}$ that only controls the magnitude of the polarization ratio, whereas the shape of the theoretical curves are governed by the magnetic order parameter of the system.\cite{Mukherjee:2012}

In the case of an out-of-plane magnetic field $H_c$, the measured $P_c(H_{c})$ increases with magnetic field for positive values of $q_c$ and exhibits a pronounced hump at the spin-flop transition at $H_{\rm{sf}}$ as seen from Fig. \ref{fig:theory1}. It can be noted that the qualitative features of the observed experimental result \cite{Zach:2011, Mukherjee:2012} is therefore reproduced for an attractive DM induced magnetoelectric coupling $q_{\alpha}P_{\alpha}^2M_iL_i$ as the polarization would be suppressed with increasing field for repulsive DM induced term. Comparing the magnitude of the polarization ratio at $q_c=0$ in Fig. \ref{fig:theory1} with the value that qualitatively matches with the experiments at q$_c=-6.58$ we find that there is a dominant contribution from DM induced magnetoelectric coupling to the polarization enhancement. On the other hand, we can also see from Fig. \ref{fig:theory1} that the coefficient of bi-quadratic magnetoelectric coupling between the magnetization and the electric polarization given by $g_c$ does not have any significant influence on the $P_c(H_{c})$.

\begin{figure}[b]
\begin{minipage}{.49\columnwidth}
\includegraphics[clip=true,width=0.99\columnwidth]{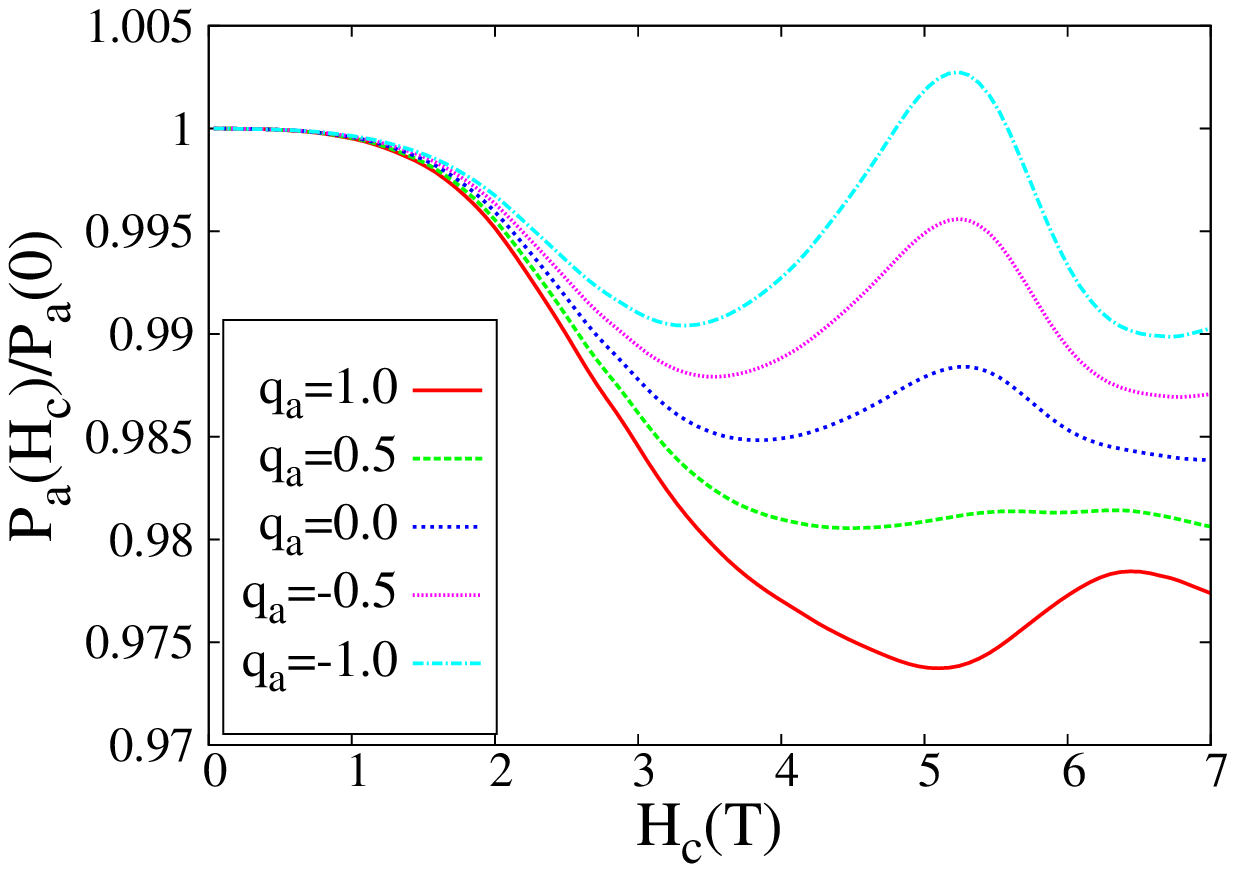}
\end{minipage}
\begin{minipage}{.49\columnwidth}
\includegraphics[clip=true,width=0.99\columnwidth]{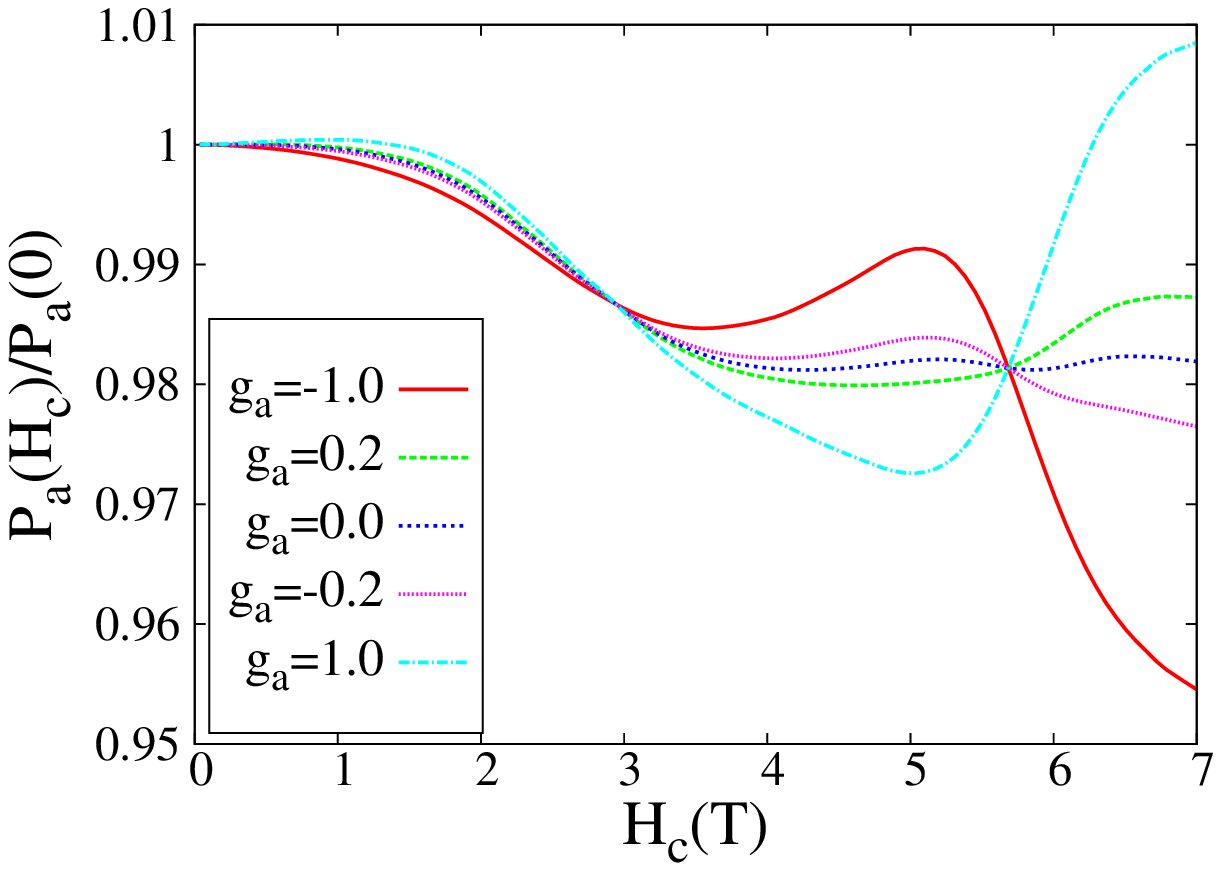}
\end{minipage}
\caption{(Color online) Theoretically calculated $P_{a}(H_{c})/P_{a}(0)$ for different values of $q_a$ and $g_a$. Parameters not mentioned in the plots have been fixed at $s_c=0.1$, $s_a=0.074$, $g_a=g_b=g_c=-0.2$, $q_{c}=-6.58$, and $q_a=q_b=0$ that explain the experimental curves at $T=5\rm{K}$.} \label{fig:theory2}
\end{figure}

For polarization values measured in the CuO$_2$ plane, the effect of the parameters on the $P_a(H_{c})$ curve is shown in Fig. \ref{fig:theory2}. As can be seen from the figure, an increase of the attractive $q_a$ term leads to the suppression of the electric polarization in contrast to the behavior of $P_c(H_{c})$ in the presence of finite $q_c$. Though the experimental curve of Ref. \onlinecite{Zach:2011} is reproduced by $q_a=0$, the qualitative features of the curve are maintained for a finite $q_a<=0$. It is also interesting to note the effect of the magnetization coupling term $g_a$ on $P_a(H_{c})$. Unlike the case for out-of-plane polarization discussed above, this term strongly influences the form of the curve and we find significant deviations from the qualitative shape of the experimental curve for different values of $g_a$. It is therefore quite remarkable that the value of $g_a$ that reproduces the experimental curve corresponds to $g_a=g_c=-0.2$ since we do not have any a priori reason for it to do so. Also note that the deviation in the polarization are much stronger above the spin flop transition. This is to be expected since the magnetization undergoes a significant enhancement above the transition.

We have observed experimentally that the magnetization shows a small upturn below the temperatures where the ferroelectric order sets in.\cite{Zach:future} This effect is in addition to the typical upturn in magnetization near the spin glass freezing temperature.\cite{Keimer:1992,Matsuda:2002} The inclusion of such a feedback effect leads to the following expression for the magnetization,
\begin{align}
M_{c}={\chi_0H_c+[1-\chi_0\sum_{\alpha}\gamma_{3\alpha}'P_{\alpha}^2(H_c)]l_{+}(H_c)\over 1+\chi_0\sum_{\alpha} \gamma_{2\alpha}'P_{\alpha}^2(H_c)},\\
M_{ab}={[1-\chi_0\sum_{\alpha}\gamma_{3\alpha}'P_{\alpha}^2(H_{ab})]l_{+}(H_{ab})\over 1+\chi_0\sum_{\alpha} \gamma_{2\alpha}'P_{\alpha}^2(H_{ab})}.
\end{align}
Note that in this expression the relative sign of the coefficients can be determined from the relations $\gamma_{3\alpha}'/\gamma_{2\alpha}'=q_\alpha/g_\alpha>0$. Therefore for values of $q_{\alpha}$ and $g_{\alpha}$ that are opposite in sign we would find a suppression in magnetization and an enhancement otherwise thus providing another experimental test to identify the relative sign of the GL coefficients. Similarly, since the magnetocapacitive effect is proportional to the polarization enhancement due to an external magnetic field the size of the jump would be indicative of the magnitude of the DM induced magnetoelectric coupling.\cite{Mukherjee:2012}

In summary, we have studied the parameter dependence of the recently discovered magnetoelectric effect in extremely underdoped La$_2$CuO$_{4+x}$ modeled by a GL free energy including a bi-quadratic magnetoelectric coupling term. Changes in this term can lead to interesting behavior of the field dependence of the polarization, particularly above the spin flop transition.

We acknowledge the financial support by the European Union through MEXT-CT-2006-039047 and EURYI research grants.  The work in Singapore was funded by The National Research Foundation.  B.M.A. acknowledges support from The Danish Council for Independent Research $|$ Natural Sciences.


\begin{thebibliography}{100}
\bibitem{coeff} The coefficients in $F_M$ are $\chi_0=5.625\times 10^{-4}\rm{cm}^3/\rm{mole}$, $C\chi_0\!=\!1.88\times 10^{-3}$, $\chi_{2D}^{-1}-C^2\chi_0-J_{\bot}=-4.4\mu \rm{eV}$, $a=2.4\times 10^{-3}/(C\chi_0)^2\rm{eV}/(\rm{emu})^2$, $b=10^{-3}/(C\chi_0)^4\rm{eV}/(\rm{emu})^2$.
\end{thebibliography}
\end{document}